\documentclass[conference]{IEEEtran}
\IEEEoverridecommandlockouts

\usepackage[pdftex]{graphicx} 
\usepackage{tikz} 
\usepackage{pgfplots} 
\pgfplotsset{compat=newest}

\usepackage{amsmath} 
\usepackage{amssymb} 
\usepackage{amsthm} 
\usepackage{thmtools} 
\usepackage{bm} 
\usepackage{csquotes} 
\usepackage{balance} 
\usepackage{cite} 

\usepackage[capitalize]{cleveref}
\crefname{equation}{\unskip}{\unskip}
\crefname{appsec}{Appendix}{Appendices}

\newtheorem{definition}{Definition}
\newtheorem{proposition}{Proposition}

\newlength\figureheight
\newlength\figurewidth
\usepackage{xcolor}

\ifCLASSOPTIONcompsoc
\usepackage[subrefformat=parens,labelformat=parens,ccaption=false,font=normalsize,labelfont=sf,textfont=sf]{subfig}
\else
\usepackage[subrefformat=parens,labelformat=parens,caption=false,font=footnotesize]{subfig}
\fi

\newcommand*\dif{\mathop{}\!\mathrm{d}}
\newcommand{\me}{\mathrm{e}}

\newcommand{\x}{\bm{x}}
\newcommand{\y}{\bm{y}}
\newcommand{\z}{\bm{z}}
\newcommand{\A}{\bm{A}}
\newcommand{\C}{\bm{C}}
\newcommand{\W}{\mathcal{W}}

\newcommand{\cred}{\sigma_{\mathsf{reduce}}}
\newcommand{\cd}{\sigma_{\mathsf{d}}}
\newcommand{\ca}{\sigma_{\mathsf{A}}}
\newcommand{\cm}{\sigma_{\mathsf{M}}}

\newcommand{\tmap}{D_{\mathsf{map}}}
\newcommand{\etmap}{\bar{D}_{\mathsf{map}}}
\newcommand{\tred}{\mathsf{D}_{\mathsf{reduce}}}

\newcommand{\emin}{\epsilon_\mathsf{min}}
\newcommand{\Pf}{P_\mathsf{f}}
\newcommand{\Pft}{P_\mathsf{f, target}}

\begin{document}

\title{A Droplet Approach Based on Raptor Codes for Distributed Computing With Straggling Servers}

\author{
  \IEEEauthorblockN{
    Albin Severinson\IEEEauthorrefmark{2},
    Alexandre Graell i Amat\IEEEauthorrefmark{3},
    Eirik Rosnes\IEEEauthorrefmark{2},
    Francisco L{\'a}zaro\IEEEauthorrefmark{4}, and
    Gianluigi Liva\IEEEauthorrefmark{4}
  }
\vspace{0.1cm}
  \IEEEauthorblockA{
    \IEEEauthorrefmark{2}Simula UiB, Bergen, Norway}

  \IEEEauthorblockA{
    \IEEEauthorrefmark{3}Department of Electrical Engineering, Chalmers
    University of Technology, Gothenburg, Sweden}

  \IEEEauthorblockA{\IEEEauthorrefmark{4}Institute of Communications and Navigation of
    DLR (German Aerospace Center), Munich, Germany}

  \thanks{This work was funded by the Research Council of Norway under
    grant 240985/F20 and the Swedish Research Council under grant
    2016-04253.}
}

\maketitle

\begin{abstract}
  We propose a coded distributed computing scheme based on Raptor
  codes to address the straggler problem. In particular, we consider a
  scheme where each server computes intermediate values, referred to
  as droplets, that are either stored locally or sent over the
  network. Once enough droplets are collected, the computation can be
  completed. Compared to previous schemes in the literature, our
  proposed scheme achieves lower computational delay when the decoding
  time is taken into account.
\end{abstract}
\IEEEpeerreviewmaketitle

\section{Introduction}
\label{sec:introduction}

Modern computing systems often consist of several thousands of servers
working in a highly coordinated manner \cite{Verma2015}. These
systems, referred to as warehouse-scale computers (WSCs)
\cite{Barroso2009}, differ from traditional datacenters in that
servers rarely have fixed roles. Instead, a cluster manager
dynamically assigns storage and computing tasks to servers
\cite{Verma2015}. This approach offers a high level of flexibility but
also poses significant challenges. For example, so-called
\emph{straggling servers}, i.e., servers that experience transient
delays, are a major issue in WSCs and may significantly slow down the
overall computation \cite{Dean2004}.

Recently, an approach based on maximum distance separable (MDS) codes
was proposed to alleviate the straggler problem for linear
computations (e.g., multiplying a matrix with a vector)
\cite{Lee2017,Li2016}. In particular, redundancy is added to the
computation in such a way that straggling servers can be treated as
erasures when decoding the final output. Any partially computed
results by the straggling servers are discarded. In \cite{Lee2017}, a
single master node is responsible for decoding the final output. A
more general framework was proposed in \cite{Li2016}, where the work
of decoding is distributed over the servers. Somewhat surprisingly,
most previous works neglect the decoding complexity of the underlying
code, which may have a significant impact on the overall computational
delay \cite{Severinson2017,Severinson2018}. For the matrix
multiplication problem, a coded scheme consisting of partitioning the
source matrix and encoding each partition separately using shorter MDS
codes was proposed in \cite{Severinson2017,Severinson2018} and shown
to significantly reduce the overall computational delay compared to
using a single MDS code when the decoding complexity is taken into
account. Furthermore, it was shown in \cite{Severinson2018} that Luby
Transform (LT) codes \cite{Luby2002} may reduce the delay further in
some cases.

Using LT codes for distributed computing has also been studied in
\cite{Keshtkarjahromi2018,Mallick2018}, where, assuming that a single
master node is responsible for decoding the output, it was shown that
these codes may bring some advantages. In \cite{Keshtkarjahromi2018},
the problem of multiplying a matrix by a vector in an
internet-of-things setting was considered. Specifically, a scheme
based on LT codes where a device may dynamically assign computing
tasks to its neighboring devices was proposed. It was shown that this
scheme achieves low delay and high resource utilization even when the
available computing resources vary over time. The scheme proposed in
\cite{Mallick2018} extends the scheme in \cite{Lee2017} by introducing
LT codes and utilizing partial computations. The authors give bounds
on the overall delay in this setting.

In this paper, we propose a coded computing scheme based on Raptor
codes \cite{Shokrollahi2011} for the problem of multiplying a matrix
by a set of vectors. In particular, we consider standardized Raptor10
(R10) codes \cite{Edmonds2017} as the underlying code. Similar to
\cite{Mallick2018}, the proposed scheme exploits partial computations,
i.e., servers compute intermediate values, referred here to as
droplets, that are either stored locally or transferred over the
network. The computation can be completed once enough droplets have
been collected. Unlike in
\cite{Lee2017,Li2016,Keshtkarjahromi2018,Mallick2018}, we take the
decoding time into account since it may contribute significantly to
the overall computational delay \cite{Severinson2018}. Furthermore,
the work of decoding the output is distributed over the servers in a
similar fashion to the scheme in \cite{Li2016}. We show that this
significantly reduces the overall computational delay compared to the
scheme in \cite{Mallick2018} when the number of servers is large, and
also outperforms other schemes in the literature. Interestingly, the
proposed scheme based on R10 codes achieves an overall computational
delay close to that of a scheme using an \emph{ideal} rateless code
with zero overhead and incurring no decoding delay. Furthermore, we
provide an analytical approximation of the expected overall
computational delay of the proposed scheme when the droplets are
computed in an optimal order. We then give a heuristic for choosing
the order in which each server computes values and show numerically
that it achieves almost identical performance to optimal ordering. We
also present an optimization problem for finding the optimal number of
servers over which the decoding of the final output should be
distributed.

\section{System Model and Preliminaries}
\label{sec:model}

We consider the distributed matrix multiplication
problem. Specifically, given an $m \times n$ matrix
$\A \in \mathbb{F}^{m \times n}_{2^u}$ and $N$ vectors
$\x_1, \ldots, \x_N \in \mathbb{F}^{n}_{2^u}$, where
$\mathbb{F}_{2^u}$ is an extension field of characteristic $2$, we
want to compute the $N$ vectors
$\bm{y}_1 = \bm{Ax}_1, \ldots, \bm{y}_N = \bm{Ax}_N$. The computation
is performed in a distributed fashion using $K$ servers,
$S_1, \ldots, S_K$. More precisely, $\A$ is split into $m/l$ disjoint
submatrices, each consisting of $l$ rows. The submatrices are then
encoded using an $(r/l, m/l)$ linear code, resulting in $r/l$ encoded
submatrices, denoted by $\C_1, \ldots, \C_{r/l}$. We refer to $l$ as
the \emph{droplet size}. Each of the $r/l$ coded submatrices is stored
at exactly one server such that each server stores $\eta m$ coded
matrix rows, for some $\frac{1}{K} \leq \eta \leq 1$. Note that,
overall, the $K$ servers store a total of $r=\eta m K$ coded rows. We
assume that $\eta$ is selected such that $\eta m$ is a multiple of
$l$. Finally, we denote by $\mathcal{C}_k$ the set of indices of the
submatrices stored by server $S_k$.

\subsection{Probabilistic Runtime Model}
\label{sec:runtime_model}

We assume that each server $S_1, \dots, S_K$ becomes available and
starts working on its assigned tasks after a random amount of time,
which is captured by the random variables $H_1, \dots, H_K$,
respectively. We assume that $H_1, \dots, H_K$ are independent and
identically distributed (i.i.d) random variables with exponential
probability density function
\begin{equation} \notag
  f_H(h) =
  \begin{cases}
    \frac{1}{\beta} \me^{-\frac{h}{\beta}} & h \geq 0 \\
    0 & h < 0
  \end{cases},
\end{equation}
where $\beta$ is used to scale the tail of the distribution. The tail
accounts for transient disturbances that are at the root of the
straggler problem. We refer to $\beta$ as the straggling parameter.
As in \cite{Mallick2018}, we assume that once a server becomes
available it carries out each of its assigned tasks in a deterministic
amount of time, denoted by $\sigma$. Let $\sigma_{\mathsf{A}}$ and
$\sigma_{\mathsf{M}}$ be the time required to compute one addition and
one multiplication, respectively, over $\mathbb{F}_{2^u}$. The
parameter $\sigma$ is then given by
$\sigma=n_{\mathsf{A}}\sigma_{\mathsf{A}}+n_{\mathsf{M}}\sigma_{\mathsf{M}}$,
where $n_{\mathsf{A}}$ and $n_{\mathsf{M}}$ are the required number of
additions and multiplications, respectively, to complete each task. As
in \cite{Edmonds2017}, we assume that $\sigma_{\mathsf{A}}$ is
$\mathcal{O}(\frac{u}{64})$ and $\sigma_{\mathsf{M}}$ is
$\mathcal{O}(u \log_2 u)$. Furthermore, we assume that the hidden
coefficients are comparable and will thus not consider them.

We denote by $H_{(i)}$, $i=1, \ldots, K$, the $i$-th order statistic,
i.e., the $i$-th smallest variable of $H_{1}, \ldots, H_{K}$.
$H_{(i)}$ is a gamma-distributed random variable with cumulative
probability distribution function
\begin{equation} \notag
    F_{H_{(i)}}(h_{(i)}) \triangleq \Pr(H_{(i)} \leq h_{(i)}) =
    \hspace{-0.3ex} \begin{cases}
      \frac{\gamma(b, ah_{(i)})}{\Gamma(b)} & \hspace{-1ex} h_{(i)}
      \hspace{-0.3ex} \geq \hspace{-0.2ex} 0 \\
      0 & \hspace{-1ex} h_{(i)} \hspace{-0.3ex} < \hspace{-0.2ex} 0
    \end{cases} \hspace{-0.3ex} ,
\end{equation}
where $\Gamma$ denotes the gamma function and $\gamma$ the lower
incomplete gamma function. The inverse scale factor $a$ and shape
parameter $b$ of the gamma distribution are computed from its mean and
variance as in \cite{Severinson2018}. The expectation of $H_{(i)}$,
i.e., the expected delay until a total of $i$ servers become
available, is \cite{Arnold2008}
\begin{equation} \notag
  \mu(K, i) \triangleq \mathbb{E}\left[H_{(i)} \right] =
  \sum_{j=K-i+1}^K \frac{\beta}{j}.
\end{equation}
Finally, we denote by $h_i$ and $h_{(i)}$ the realizations of $H_i$
and $H_{(i)}$, $i=1, \dots, K$, respectively.

\subsection{Distributed Computing Model}
\label{sec:computing_model}

We consider the coded computing framework introduced in \cite{Li2016},
which extends the MapReduce framework \cite{Dean2004}. The overall
computation proceeds in two phases, the \textit{map-shuffle} phase and
the \textit{reduce} phase, which are augmented to make use of the
coded scheme proposed in \cite{Mallick2018} to alleviate the straggler
problem.  We assume that the input vectors $\x_1, \ldots, \x_N$ are
known to all servers at the start of the computation.

\subsubsection{Map-Shuffle Phase}

The servers compute coded intermediate values (droplets) which are
later used to obtain the vectors $\y_1,\ldots,\y_N$. Each droplet is
the product between a submatrix stored by the server and an input
vector $\x_1, \ldots, \x_N$. The responsibility for decoding each of
the vectors $\y_1,\ldots,\y_N$ is assigned to one of the $K$
servers. The computed droplets are then transferred over the network
to the server responsible for decoding the corresponding output
vector. We assume that the channel is error-free and that all
transfers are unicast. The map-shuffle phase ends when all output
vectors $\y_1, \ldots, \y_N$ can be decoded with high probability (see
\cref{sec:code_design}). At this point the computation enters the
reduce phase. We denote the delay of the map-shuffle phase by $\tmap$
and its expectation by $\etmap$.

\subsubsection{Reduce Phase}

The vectors $\y_1,\ldots,\y_N$ are computed from the intermediate
values. More specifically, each server uses the droplets computed
locally or received over the network to decode the output vectors it
has been assigned. Denote by $\cred$ the time required for one server
to decode one output vector. The computational delay of the reduce
phase, denoted by $\tred$, is deterministic and is given by
$\tred = \frac{N}{q} \cred$, where $q$ denotes the number of servers
used in the reduce phase.

\begin{definition}
\label{def:D}
The \textit{overall computational delay}, $D$, is the sum of the
map-shuffle and reduce phase delays, i.e.,
\begin{equation}
  \notag
  D = \tmap + \tred \;\;\;\;\text{and}\;\;\;\; \bar{D} \triangleq \mathbb{E}[D] = \etmap + \tred.
\end{equation}
\end{definition}

\subsection{Raptor Codes}
\label{sec:raptor_codes}

Raptor codes \cite{Shokrollahi2011} are built from the serial
concatenation of an outer linear block code with an inner LT
code. Raptor codes not only outperform LT codes in terms of
probability of decoding failure but also exhibit a lower encoding and
decoding complexity. Here we consider R10 codes, which are binary
codes whose outer code is obtained as the serial concatenation of a
low-density parity-check code with a high-density parity-check (HDPC)
code \cite{rfc5053}.  R10 codes are tailored to an efficient maximum
likelihood decoding algorithm known as \emph{inactivation decoding}
\cite{Shokrollahi2011}. In particular, we consider R10 codes in their
nonsystematic form.

\section{Proposed Coded Computing Scheme}

\begin{figure}[t]
\centering
  \hspace{-2ex} \resizebox{1.03\columnwidth}{!}{
    \includegraphics{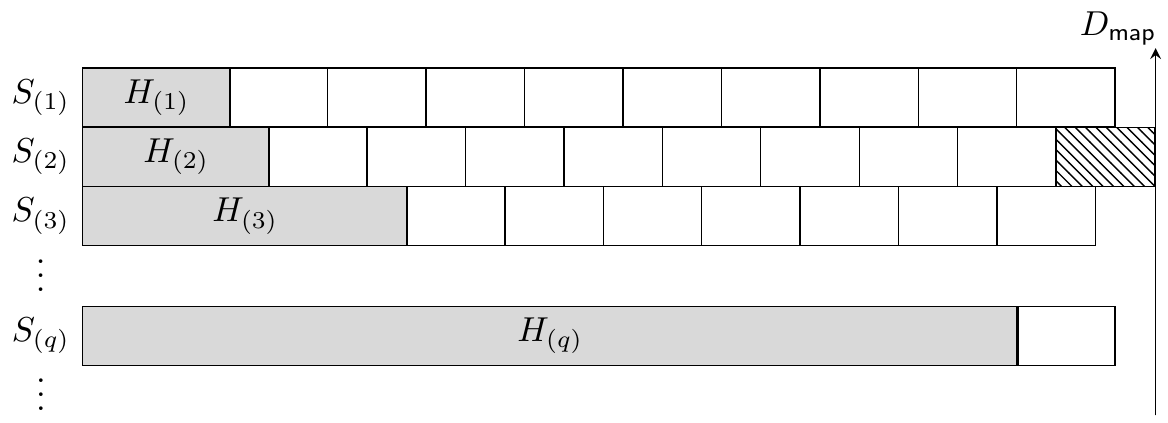}
  }
  \caption{Map-shuffle phase computation. Each server $S_{(k)}$,
    $k=1, \ldots, K$, computes droplets, illustrated by white squares,
    after an initial time $H_{(k)}$. The map-shuffle phase ends once
    enough droplets are collected and server $S_{(q)}$ has become
    available. It incurs a delay $\tmap$. We depict the final droplet
    that is computed with a hash pattern.}
  \label{fig:overview}
\end{figure}%

In this section, we introduce the proposed coded computing scheme. The
main idea is that each server computes multiple intermediate values.
More specifically, each server $S_k$, $k=1, \ldots, K$, computes
droplets $\z_j^{(i)} = \C_i \x_j$ by multiplying the coded submatrices
$\C_i$, $i\in \mathcal{C}_k$, it stores locally with the $N$ input
vectors $\x_1, \ldots, \x_N$. The indices $i \in \mathcal{C}_k$ and
$j \in \{1, \ldots, N\}$ should be carefully chosen to minimize the
computational delay. We consider this in \cref{sec:droplet_order}.
The time required for a server to compute a droplet, denoted by $\cd$,
is
$\cd = l \left( (n-1)\sigma_{\mathsf{A}} + n\sigma_{\mathsf{M}}
\right)$ since it requires computing $l$ inner products, each
requiring $n-1$ additions and $n$ multiplications.

Denote by $S_{(1)}$ the first server to become available, and
similarly denote by $S_{(k)}$, $k=1, \dots, K$, the $k$-th server to
become available. We assume that server $S_{(k)}$ computes droplets at
a constant rate after a delay $H_{(k)}$. For example, server
$S_{(k)}$ computes $p$ droplets after a total delay of
$H_{(k)} + p\cd$. This process is depicted in \cref{fig:overview}. We
evenly and randomly split the indices of the $N$ output vectors
$\y_1,\ldots,\y_N$ into $q\leq K$ disjoint sets $\W_1, \ldots,
\W_q$. Each of the $q$ fastest servers $S_{(k)}$, $k=1, \ldots, q$, is
responsible for decoding the $N/q$ output vectors with indices in
$\W_k$. Furthermore, we denote by $\tilde{\W}_k$ the set containing
the indices of the vectors that server $S_{(k)}$ is not yet able to
decode due to an insufficient number of droplets. At the start of the
map-shuffle phase, $\tilde{\W}_k=\W_k$. The map-shuffle phase ends
when servers $S_{(1)}, \ldots, S_{(q)}$ have collected enough droplets
to decode the output vectors they are responsible for, i.e., when
$\left\vert \tilde{\W}_k \right\vert = 0$, $k=1, \ldots, q$. At this
point servers $S_{(1)}, \ldots, S_{(q)}$ simultaneously enter the
reduce phase. The remaining $K-q$ servers are unused for the rest of
the computation. A strategy for choosing $q$ to minimize the expected
computational delay, $\bar{D}$, is discussed in
\cref{sec:straggler_mitigation}.

\subsection{Droplet Order}
\label{sec:droplet_order}

For each droplet $\z_j^{(i)}$ computed by server $S_k$ in the
map-shuffle phase, the server has to choose the indices
$i \in \mathcal{C}_k$ and $j \in \{1, \ldots, N\}$ the droplet is
computed from. Furthermore, the choice of $i$ and $j$ may have a large
impact on the computational delay. In particular, if $j$ is chosen
such that it is not needed to decode any of the output vectors, i.e.,
$j$ is not in any of the sets $\tilde{\W}_k$, $k=1, \ldots, q$, the
resulting droplet is effectively wasted. Hence, $i$ and $j$ should be
carefully chosen. We consider two scenarios. In the first scenario,
$i$ and $j$ are chosen optimally, i.e., all servers have perfect
knowledge of $\tilde{\W}_1, \ldots, \tilde{\W}_q$. This gives a lower
bound on the achievable computational delay. In a second, more
practical scenario, $i$ and $j$ are chosen in a round-robin
fashion. Specifically, for each server $S_k$ we generate a number $j$
from $\{1, \ldots, N\}$ uniformly at random. Next, for each droplet
$\z_j^{(i)}$ computed by server $S_k$ we let $j = j+1 \bmod N$. We
remark that the optimal order requires each server to have global
knowledge of all previously computed droplets over all servers,
whereas the round-robin strategy only requires each server to have
knowledge of the droplets it has computed locally. In
\cref{sec:numerical_results}, we show numerically that the round-robin
strategy achieves almost identical performance to the optimal
strategy, the latter being infeasible in practice. In both cases we
assume that the same pair of indices $i, j$ is never chosen
twice. Since each submatrix $\C_i$ is stored at exactly one server,
this does not require any additional synchronization between
servers. A server that has exhausted all possible combinations of $i$
and $j$ halts and performs no further computations in the map-shuffle
phase.

\subsection{Code Design}
\label{sec:code_design}

The decoding complexity and failure probability of Raptor codes depend
on the number of droplets available to the decoder,
$\frac{m}{l}(1+\epsilon)$, for some $\epsilon \geq 0$. We refer to
$\epsilon$ as the overhead. Furthermore, we denote by $\Pf(\epsilon)$
the decoding failure probability when the overhead is $\epsilon$. In
general, increasing $\epsilon$ reduces the probability of decoding
failure $\Pf(\epsilon)$ and the decoding complexity, leading to a
lower decoding time $\cred$. For example, the decoding failure
probability for R10 codes roughly halves with every additional droplet
available when the number of source symbols is close to $1000$
\cite{Shokrollahi2011}. However, a larger overhead $\epsilon$ also
increases the computational delay due to computing the required
droplets in the map-shuffle phase. We thus need to balance the
computational delay of the reduce phase against that of the
map-shuffle phase to achieve a low overall computational delay.

We denote by $\emin$ the minimum overhead before decoding is
attempted.  R10 codes are fully specified, hence the only free
parameter is $\emin$. In \cite{Shokrollahi2011}, it is observed that
the decoding complexity of Raptor codes drops sharply when the number
of droplets available to the decoder is increased to be slightly
larger than the number of HDPC symbols.  Hence, we choose the minimum
overhead $\emin$ such that the number of droplets available to the
decoder is close to the number of source droplets $m/l$ plus twice the
number of HDPC symbols. For comparison purposes, in
\cref{sec:numerical_results} we also consider LT codes with a robust
Soliton distribution \cite{Luby2002}, whose parameters are optimized
as described in \cite{Severinson2018}. In particular, we choose a
minimum overhead $\emin$ and a target failure probability $\Pft$ and
optimize the parameters of the distribution to minimize the decoding
complexity under the constraint
$\Pft \approx \Pf(\epsilon_\mathsf{min})$. Note that the overhead
$\epsilon$ required for decoding may be larger than $\emin$. We take
this into account by simulating the overhead needed given that
decoding failed at an overhead of $\emin$.

\section{Computational Delay Analysis}

In this section, we analyze the computational delay of the proposed
coded computing scheme and provide an approximation of $\etmap$. Let
$V_p$ be the random variable associated with the time until $p$
droplets are computed over $K$ servers, where we assume that $p$ is
chosen such that decoding succeeds with high probability, and
$\bar{V}_p$ its expectation, $\bar{V}_p \triangleq
\mathbb{E}[V_p]$. Then, $\tmap = \max(V_p, H_{(q)})$. For the
analysis, we assume that each server is always able to compute
droplets needed by some server until the end of the map-shuffle
phase. This assumption is valid if the code rate $m/r$ is low
enough. Furthermore, we assume that the droplet order is optimal (see
\cref{sec:droplet_order}). Finally, we explain how to choose the
number of servers $q$ to split the output vectors over to minimize the
expected computational delay.

Denote by $P_t$ the number of droplets computed over $K$ servers at
time $t$.
\begin{proposition} The expectation of $P_t$ is
  \begin{equation} 
  \label{eq:expectation}
    \bar{P}_t \triangleq \mathbb{E}\left[ P_t \right] =
    K \int_0^t  \left\lfloor
      {\frac{t-h}{\cd}} \right\rfloor \frac{1}{\beta} {\rm{e}}^{-\frac{h}{\beta}} \dif h.
\end{equation}
\end{proposition}


Using the fact that $x - 1 \leq \lfloor x \rfloor \leq x$ in
\eqref{eq:expectation} and computing the resulting integrals,
$\bar{P}_t$ can be lower and upperbounded as
\begin{multline}
  \label{eq:dt}
  K \left( \frac{(\beta + \cd) {\rm{e}}^{-\frac{t}{\beta}}}{\cd} +
    \frac{t}{\cd} - \frac{\beta}{\cd} -1 \right) \leq \bar{P}_t \\ \leq
  K \left( \frac{\beta {\rm{e}}^{-\frac{t}{\beta}}}{\cd} +
    \frac{t}{\cd} - \frac{\beta}{\cd}  \right).
\end{multline}

Let $\sigma_{\bar{P}}$ denote the time at which an average number of
$\bar{P}$ droplets have been computed over $K$ servers. By inverting
the upper and lower bounds on $\bar{P}_t$ in \cref{eq:dt},
$\sigma_{\bar{P}}$ can be bounded as
\begin{multline*}
  \sigma^{\mathsf{L}}_{\bar{P}} \triangleq \beta + \frac{\bar{P} \cd}{K} +
  \beta W_0 \left(- {\rm{e}}^{-\frac{\bar{P}\cd}{K\beta}-1} \right)
  \leq \sigma_{\bar{P}} \\ \leq
  \beta + \cd + \frac{\bar{P} \cd}{K} + \beta W_0 \left(-\frac{
      {\rm{e}}^{-\frac{K(\beta+\cd)+\bar{P}\cd}{K\beta}}
      (\beta+\cd)}{\beta} \right)
  \triangleq  \sigma^{\mathsf{U}}_{\bar{P}},
\end{multline*}
where $W_0(\cdot)$ is the principal branch of the Lambert W function,
i.e., $W_0(x)$ is the solution of $x=z {\rm{e}}^z$.

Now, let $G_t$ be the random variable associated with the number of
servers that are available at time $t$. We provide the following
heuristic approximation of $\etmap$,
\begin{equation} \label{eq:edmap}
  \etmap \approx
  \bar{V}_p + \sum_{j=1}^{q-1} \Pr(G_t=j) \mu(K-j, q-j),
\end{equation}
where the summation accounts for the delay due to waiting for server
$S_{(q)}$. We have numerically verified that the approximation
holds. Furthermore, we have observed that
$\bar{V}_p \approx \sigma_{p}$ and
$\sigma_{p} \approx \frac{1}{2}\left( \sigma^{\mathsf{L}}_{p}+
  \sigma^{\mathsf{U}}_{p}\right)$. Finally, assuming that decoding is
possible with $p$ droplets, the expected overall computational delay
is
  \begin{align}
    \bar{D}
    \approx \frac{N}{q} \cred + \bar{V}_p + \sum_{j=1}^{q-1} \Pr(G_t=j) \mu(K-j, q-j).
    \label{eq:dest}
  \end{align}

\subsection{Straggler Mitigation}
\label{sec:straggler_mitigation}

The map-shuffle phase ends when all output vectors can be decoded and
when the servers $S_{(1)}, \ldots, S_{(q)}$ are available, i.e.,
$\tmap = \max(V_p, H_{(q)})$. Since $\Pr(H_{(q)} > V_p)$ is always
nonzero, choosing a small $q$ lowers the expected delay of the map
phase. On the other hand, choosing a large $q$ reduces the delay of
the reduce phase $\tred = \frac{N}{q} \cred$, as the decoding is
distributed over more servers. Thus, we need to balance the delay of
the map-shuffle and reduce phases by choosing $q$ carefully. In
particular, we optimize the value of $q$ to minimize the overall
computational delay in \cref{eq:dest}, where we use the approximation
$\bar{V}_p \approx \sigma_{p} \approx \frac{1}{2}\left(
  \sigma^{\mathsf{L}}_{p}+ \sigma^{\mathsf{U}}_{p}\right)$. We remark
that \cref{eq:dest} as a function of $q$ is convex as it is the sum of
the approximation of $\etmap$ in \cref{eq:edmap} and $\tred$, which are
strictly increasing and decreasing, respectively, in $q$ for
$\cred > 0$. For $\cred = 0$, \cref{eq:dest} is minimized for $q=1$.

\section{Numerical Results}
\label{sec:numerical_results}

\begin{figure}[t]
  \vspace{-2ex}
  \hspace{-2ex} \resizebox{1.03\columnwidth}{!}{
    \definecolor{orange}{rgb}{1,0.5,0}
\definecolor{commline}{rgb}{0.9,0.9,0.9}





\begin{tikzpicture}

\definecolor{color0}{rgb}{0.12156862745098,0.466666666666667,0.705882352941177}
\definecolor{color1}{rgb}{1,0.498039215686275,0.0549019607843137}
\definecolor{color2}{rgb}{0.75,0,0.75}
\definecolor{color3}{rgb}{0,0.75,0.75}
\definecolor{mycolor2}{rgb}{0.46600,0.67400,0.18800}%

\begin{axis}[
xlabel={\large{number of servers ($K$)}},
ylabel={\large{computational delay ($\bar{D}$)}},
xmin=0, xmax=600,
ymin=0.2, ymax=0.7,
width=11.5cm,
height=10.5cm,
tick align=outside,
tick pos=left,
xmajorgrids,
x grid style={white!69.01960784313725!black},
ymajorgrids,
y grid style={white!69.01960784313725!black},
legend style={draw=white!80.0!black},
legend cell align={left},
]
\addplot [line width=1.5pt, blue, mark=*, mark size=3, mark repeat=0, mark options={solid, line width = 0.5pt, blue, fill=white}]
table {%
20 0.341133761732947
21 0.337899606975408
22 0.335553895605125
24 0.330607066770735
42 0.301824650241078
49 0.294585252066369
65 0.282410659189506
68 0.280650643397005
73 0.277687231245031
75 0.276609248624348
77 0.275567433639825
94 0.267977716940706
104 0.264247557652062
121 0.258799138710597
152 0.251112812243698
169 0.247636686198925
179 0.245810993045416
191 0.243611729395425
204 0.241614114001368
219 0.239443281256972
236 0.237246844400203
237 0.237099960472728
257 0.234743187401666
281 0.232199191607426
282 0.232117566738782
311 0.229374185681408
312 0.229302392190274
347 0.226399336962794
625 0.211644973547967
};
\addlegendentry{R10}

\addplot [line width=1.5pt, brown, mark=x, mark size=3, mark repeat=0, mark options={solid,rotate=180}]
table {%
20 0.343476967722174
21 0.339563272412246
22 0.335239239056635
24 0.331504173947373
42 0.299199496451249
49 0.293304924575617
65 0.279730362424745
68 0.277912377951078
73 0.279101894125219
75 0.276428941935496
77 0.27346305892159
94 0.266715341118142
104 0.263403938367242
121 0.258097035316774
152 0.249961237254073
169 0.24761714351508
179 0.245925806793885
191 0.243658555947435
204 0.241819677931181
219 0.239561234770181
236 0.237280186539436
237 0.237657831705914
257 0.23543235021988
281 0.232627740395482
282 0.232310211531787
311 0.229504757695375
312 0.229438791034969
347 0.226235389437805
625 0.211579179175841
};
\addlegendentry{R10 sim. (opt.)}

\addplot [line width=1.5pt, mycolor2, dashed]
table {%
20 0.345119341902987
21 0.341705453301911
22 0.338243792478013
24 0.332404695262625
42 0.301613614665286
49 0.294313024230398
65 0.282819958598537
68 0.281491649013562
73 0.279293325018555
75 0.278171265346596
77 0.277286975144621
94 0.268968993967415
104 0.265788025834878
121 0.260104977740153
152 0.252432435662002
169 0.249008960619993
179 0.246393275659728
191 0.244382152958636
204 0.242581147241191
219 0.241022734083851
236 0.238245893852372
237 0.238179742993839
257 0.235892266137938
281 0.233410229863743
282 0.233096499852158
311 0.230583491131049
312 0.230524798443291
347 0.227452199884216
625 0.213709072072546
};
\addlegendentry{R10 sim. (rr)}

\addplot [line width=1.5pt, color2, mark=diamond*, mark size=3, mark repeat=0, mark options={solid, line width = 0.5pt, color2, fill=white}]
table {%
20 0.535973822439165
21 0.524910565259575
22 0.527207641551156
24 0.519435693020142
42 0.47421421892079
49 0.460547075034088
65 0.44204670180318
68 0.440946522705099
73 0.434830356822203
75 0.433180642054706
77 0.431585311487819
94 0.421035373344423
104 0.415174713721178
121 0.405789060471529
152 0.394537980419327
169 0.38850971538178
179 0.385676698306805
191 0.379399350839592
204 0.376763687334966
219 0.372967416232525
236 0.36995354519504
237 0.369348168057159
257 0.365705738024685
281 0.361771399749585
282 0.361955184658976
311 0.357399546261256
312 0.357565328751894
347 0.352792656301556
625 0.329903319287754
};
\addlegendentry{LT};

\addplot [line width=1.5pt, black]
table {%
20 0.328939404556203
21 0.32616469192789
22 0.323559397659373
24 0.318789571641752
42 0.291036113592752
49 0.284187537635838
65 0.272411959438014
68 0.27061895786839
73 0.267845566398283
75 0.266803583425022
77 0.265796615902149
94 0.258399016081925
104 0.254802188413952
121 0.24959605245215
152 0.242136934696273
169 0.238817699589263
179 0.237055223898002
191 0.235097095021193
204 0.233141919500934
219 0.231070886612406
236 0.228927787573662
237 0.228807727981473
257 0.226531715821319
281 0.224075043596202
282 0.22397836566429
311 0.221347175566056
312 0.221261904018951
347 0.218474798342238
625 0.204230997875334
};
\addlegendentry{Ideal rateless}

\addplot [line width=1.5pt, color3, mark=triangle*, mark size=3, mark repeat=0, mark options={solid, line width = 0.5pt, color3, fill=white}]
table {%
20 0.377890455188037
21 0.375499465388518
22 0.370292954210604
24 0.364919038556578
42 0.332483899468766
49 0.323938402664868
65 0.31084051185197
68 0.30916473615401
73 0.305261286378359
75 0.303527926452253
77 0.302705726330042
94 0.294859939297214
104 0.290424929342616
121 0.28486653782197
152 0.275668984320275
169 0.271862354376742
179 0.269604347834221
191 0.266927050115157
204 0.264512353646331
219 0.261915230316715
236 0.259376015428929
237 0.258652030037637
257 0.256995214601517
281 0.253912941121932
282 0.253525964139118
311 0.250613184190023
312 0.250597125585584
347 0.247310633904824
625 0.231988011995928
};
\addlegendentry{BDC \cite{Severinson2017,Severinson2018}}

\addplot [line width=1.5pt, red, mark=square*, mark size=2, mark repeat=0, mark options={solid, line width = 0.5pt, red, fill=white}]
table {%
20 0.333603480566641
21 0.330789424959511
22 0.328147190255628
24 0.323309732477725
42 0.295162754587278
49 0.288217071990938
65 0.276274526276119
68 0.274456101561387
73 0.271643385821808
75 0.270586628668802
77 0.269565383315118
94 0.299223998239635
104 0.325433599583239
121 0.369459539848149
152 0.448260368917199
169 0.49077904248358
179 0.515588163217956
191 0.528783338616144
204 0.559657836211394
219 0.59503853240376
236 0.634844412907965
237 0.637176751884938
257 0.683622638056548
281 0.738883573041627
282 0.741175565364016
311 0.807303020929685
312 0.809571959915116
347 0.888546826615729
625 1.49320951876599
};
\addlegendentry{Cent.\ R10}

\addplot [line width=1.5pt, green!50.0!black, dotted]
table {%
20 0.789184828971169
21 0.778312008306122
22 0.768375271256837
24 0.760973402135082
42 0.695483170837157
49 0.677730484170872
65 0.652508380006838
68 0.64819236488646
73 0.639578296381343
75 0.638081988314054
77 0.636622028110808
94 0.617423067533064
104 0.610279434851812
121 0.596728260273585
152 0.580039048586691
169 0.571383940440406
179 0.567945286543405
191 0.563277011781605
204 0.558323452184084
219 0.553433111986253
236 0.548641268270884
237 0.548091640255395
257 0.542964113679071
281 0.537155574363522
282 0.536715873471321
311 0.530726091826935
312 0.530329436768818
347 0.523962895077288
625 0.490669572856805
};
\addlegendentry{MDS \cite{Lee2017}}

\end{axis}

\end{tikzpicture}%
  }
  \caption{Performance dependence on system size for
    $mnN/K \approx 10^7$, $n=m/1000$, $N=10K$, $m/r=1/3$,
    $m/l \approx 1024$, and $\beta=\sigma_{\mathsf{K}}$.}
    \label{fig:workload}
\end{figure}
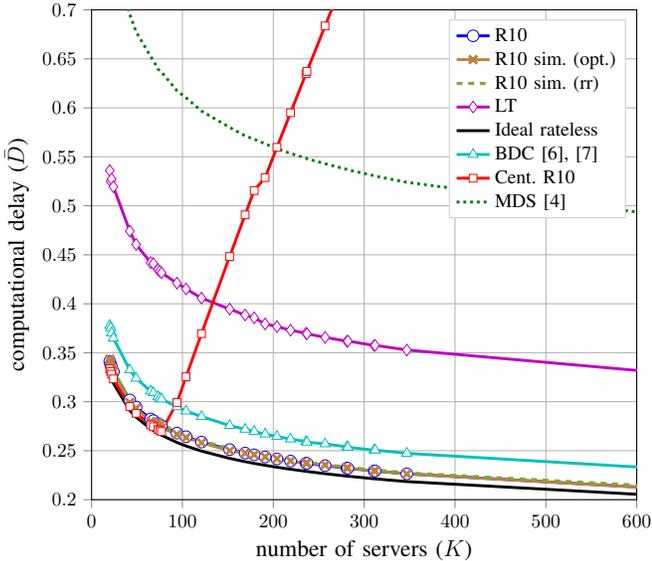

In \cref{fig:workload}, we give the expected computational delay of
the proposed scheme, normalized by that of the uncoded scheme, as a
function of the system size. In particular, we fix the code rate to
$m/r=1/3$ and the problem size divided by the number of servers to
$mnN/K=10^7$ ($\pm 10$\% to find valid parameters) and scale the
system size with $K$. Motivated by machine learning applications,
where the number of rows and columns often represent the number of
samples and features, respectively, we set $m=1000n$. We also set
$N=10K$. Since R10 codes are optimized for code lengths close to
$1024$ \cite{rfc5053}, we choose the droplet size $l$ such that
$900 < m/l < 1100$ (the interval is required to find valid
parameters). The overhead is $2$\% and $30$\% for R10 and LT codes,
respectively. Finally, the straggling parameter $\beta$ is equal to
the total time required to compute the multiplications
$\A \x_1, \dots, \A \x_N$ divided by the number of servers, i.e.,
$\beta=\sigma_{\mathsf{K}} = (m(n-1)\ca + mn\cm)N/K$.

In the figure, we plot the overall computational delay given by
\eqref{eq:dest} using the approximation
$\bar{V}_p \approx \sigma_{p} \approx \frac{1}{2}\left(
  \sigma^{\mathsf{L}}_{p}+ \sigma^{\mathsf{U}}_{p}\right)$ for the
proposed scheme with an underlying R10 code (blue line with circle
markers) and LT code (magenta line with diamond markers), and for the
scheme assuming an ideal rateless code (black solid line). We also
show simulated performance for the R10-based scheme with optimal
droplet ordering and with a round-robin (rr) ordering. We observe that
the round-robin strategy achieves a computational delay within $1$\%
of that of the optimal strategy. Furthermore, \cref{eq:dest}
accurately predicts the overall computational delay with an error of
at most about $1$\% compared to both the optimal and the round-robin
ordering. The proposed scheme with R10 codes achieves a significantly
lower delay than the scheme with LT codes. Interestingly, the delay
for the scheme based on R10 codes is very close (at most $3.7$\%
higher) to that of an ideal rateless code.

For comparison purposes, we also plot in the figure the delay of the
block-diagonal coding (BDC) scheme in
\cite{Severinson2017,Severinson2018}, the MDS coding scheme proposed
in \cite{Lee2017} that does not utilize partial computations, and the
scheme proposed in \cite{Mallick2018} (augmented with R10 codes). We
refer to it as the centralized R10 (cent.\ R10) scheme, since a
central master node is responsible for decoding all output
vectors. For small $K$, the delay is limited by the time needed to
compute droplets.  However, for $K\gtrsim 90$ the master node of the
centralized scheme can no longer decode the output vectors quickly
enough, causing a high overall computational delay.  Thus, for
$K\gtrsim 90$ the scheme in \cite{Mallick2018} (now with R10 codes),
incurs a delay significantly higher than that of the proposed
scheme. The proposed scheme also yields a significantly lower
computational delay than that of the scheme in
\cite{Lee2017}. Finally, the delay of the BDC scheme in
\cite{Severinson2017,Severinson2018} is about $10$\% higher compared
to the proposed scheme based on R10 codes.

\begin{figure}[t]
  \vspace{-2ex}
  \hspace{-2ex} \resizebox{1.01\columnwidth}{!}{
    \definecolor{orange}{rgb}{1,0.5,0}
\definecolor{commline}{rgb}{0.9,0.9,0.9}





\begin{tikzpicture}

\definecolor{color0}{rgb}{0.75,0,0.75}
\definecolor{color1}{rgb}{0,0.75,0.75}

\begin{axis}[
xlabel={\large{$\beta/\sigma_{\mathsf{K}}$}},
ylabel={\large{computational delay ($\bar{D}$)}},
xmin=1, xmax=5,
ymin=0, ymax=1.6,
width=11.5cm,
height=10.5cm,
tick align=outside,
tick pos=left,
xmajorgrids,
x grid style={white!69.01960784313725!black},
ymajorgrids,
y grid style={white!69.01960784313725!black},
legend style={draw=white!80.0!black},
legend cell align={left}
]

\addplot [line width=1.5pt, blue, mark=*, mark size=3, mark repeat=0, mark options={solid, line width = 0.5pt, blue, fill=white}]
table {%
1 0.211644973547967
1.21052631578947 0.190709516170834
1.42105263157895 0.17470912297606
1.63157894736842 0.161991783654589
1.84210526315789 0.151584162305832
2.05263157894737 0.142871775223933
2.26315789473684 0.135445617449945
2.47368421052632 0.129021814255238
2.68421052631579 0.123396527667395
2.89473684210526 0.118419210220351
3.10526315789474 0.113976003491851
3.31578947368421 0.109979034840751
3.52631578947368 0.106359309038541
3.73684210526316 0.103062050002202
3.94736842105263 0.100044624909381
4.15789473684211 0.0972787672713907
4.36842105263158 0.0947579818689253
4.57894736842105 0.0924769045552415
4.78947368421053 0.0905015043763132
5 0.0888709854050875
};
\addlegendentry{R10};

\addplot [line width=1.5pt, color0, mark=diamond*, mark size=3, mark repeat=0, mark options={solid, line width = 0.5pt, color0, fill=white}]
table {%
1 0.329903319287754
1.21052631578947 0.291596285512755
1.42105263157895 0.262926535080656
1.63157894736842 0.240546246577663
1.84210526315789 0.222516211210562
2.05263157894737 0.207631016276722
2.26315789473684 0.195099588494191
2.47368421052632 0.18437991765708
2.68421052631579 0.175087393066487
2.89473684210526 0.16694103878349
3.10526315789474 0.159730489741415
3.31578947368421 0.153294907428416
3.52631578947368 0.147509067734277
3.73684210526316 0.142273916877587
3.94736842105263 0.137509999884253
4.15789473684211 0.133152787661758
4.36842105263158 0.12914929328942
4.57894736842105 0.125455607743755
4.78947368421053 0.122035211413429
5 0.118858214885101
};
\addlegendentry{LT};

\addplot [line width=1.5pt, black]
table {%
1 0.204230997875334
1.21052631578947 0.184361557049961
1.42105263157895 0.169140070771083
1.63157894736842 0.157017871404009
1.84210526315789 0.147080516692034
2.05263157894737 0.138749581398041
2.26315789473684 0.131639393562904
2.47368421052632 0.125481856316246
2.68421052631579 0.120084190452916
2.89473684210526 0.115303834275691
3.10526315789474 0.111032843134912
3.31578947368421 0.107187818278014
3.52631578947368 0.103703192098992
3.73684210526316 0.100526625125201
3.94736842105263 0.0976157734121462
4.15789473684211 0.0949359696210224
4.36842105263158 0.0924585279438119
4.57894736842105 0.0901594840787761
4.78947368421053 0.0880186443820596
5 0.0860188585067178
};
\addlegendentry{Ideal rateless};

\addplot [line width=1.5pt, color1, mark=triangle*, mark size=3, mark repeat=0, mark options={solid, line width = 0.5pt, color1, fill=white}]
table {%
1 0.232713289341297
1.21052631578947 0.208445447909992
1.42105263157895 0.190114025558386
1.63157894736842 0.175574124274033
1.84210526315789 0.163784515026162
2.05263157894737 0.153851933038012
2.26315789473684 0.145426482532427
2.47368421052632 0.13820007236179
2.68421052631579 0.131266041782592
2.89473684210526 0.125746697068748
3.10526315789474 0.120758385537925
3.31578947368421 0.116306267150784
3.52631578947368 0.11237892014154
3.73684210526316 0.108840266843121
3.94736842105263 0.105658100774927
4.1578947368421 0.102870966093156
4.36842105263158 0.100302224757131
4.57894736842105 0.0980174669001124
4.78947368421053 0.0959173944802677
5 0.0939594942796864
};
\addlegendentry{BDC \cite{Severinson2017,Severinson2018}};

\addplot [line width=1.5pt, red, mark=square*, mark size=2, mark repeat=0, mark options={solid, line width = 0.5pt, red, fill=white}]
table {%
1 1.49320951876599
1.21052631578947 1.25794913838274
1.42105263157895 1.08675842689603
1.63157894736842 0.956599943466174
1.84210526315789 0.854299105742736
2.05263157894737 0.771776295750381
2.26315789473684 0.703801011383603
2.47368421052632 0.646837738637141
2.68421052631579 0.598410803173367
2.89473684210526 0.556734916621397
3.10526315789474 0.520490198864755
3.31578947368421 0.488679795000803
3.52631578947368 0.460536750309551
3.73684210526316 0.435461345755239
3.94736842105263 0.412977870874283
4.15789473684211 0.392704145781154
4.36842105263158 0.37432961096173
4.57894736842105 0.357599300646768
4.78947368421053 0.342301935463357
5 0.328260949866988
};
\addlegendentry{Cent.\ R10};

\addplot [line width=1.5pt, green!50.0!black, dotted]
table {%
1 0.490669572856805
1.21052631578947 0.440928609186253
1.42105263157895 0.40473644711725
1.63157894736842 0.377220871105704
1.84210526315789 0.355595711636706
2.05263157894737 0.338152361667054
2.26315789473684 0.323784765158471
2.47368421052632 0.311745320849244
2.68421052631579 0.30151055395336
2.89473684210526 0.292702982415729
3.10526315789474 0.285043538996218
3.31578947368421 0.278321461163812
3.52631578947368 0.272374598217194
3.73684210526316 0.267076160130672
3.94736842105263 0.26232557688062
4.15789473684211 0.258042053805888
4.36842105263158 0.254159938745523
4.57894736842105 0.250625333316018
4.78947368421053 0.247393575232619
5 0.244427341194633
};
\addlegendentry{MDS \cite{Lee2017}};

\end{axis}
\end{tikzpicture}%
  }
  \caption{Performance dependence on the straggling parameter $\beta$
    for $K=625$, $m=33333$, $n=33$, $N=6250$, $l=32$, and $m/r=1/3$.} 
  \label{fig:straggling}
\end{figure}
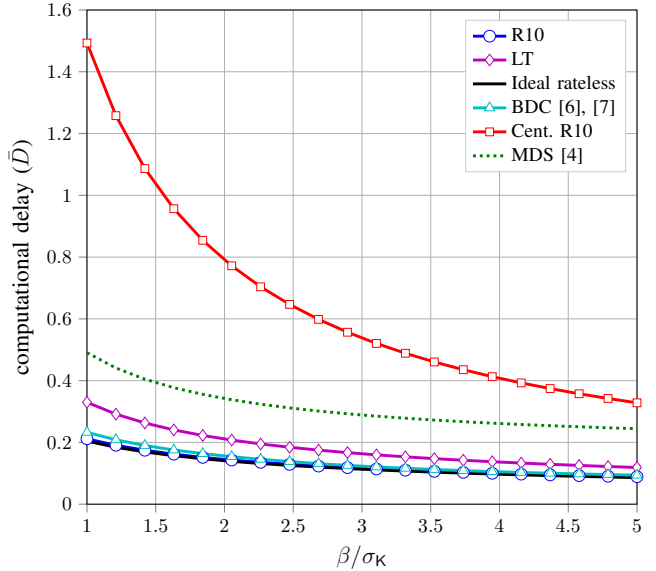

In \cref{fig:straggling}, we give the expected computational delay as a
function of the straggling parameter $\beta$ for $K=625$, $m=33333$,
$n=33$, $N=6250$, $l=32$, and $m/r=1/3$. Since $l$ is not a divisor of
$m$, $\bm A$ is zero-padded with $11$ all-zero rows. The performance
of the centralized scheme approaches that of our scheme as $\beta$
grows since the average rate at which droplets are computed decreases
with $\beta$. The scheme based on R10 codes operates close to an ideal
rateless code for all values of $\beta$ considered.


\section{Conclusion}
\label{sec:conclusion}

We introduced a coded computing scheme based on Raptor codes for
distributed matrix multiplication where each server computes several
intermediate values and where the work of decoding the output is
distributed among servers. Compared to previous schemes, the proposed scheme yields
significantly lower computational delay when the number of servers is
large. For instance, the delay is less than half when the number of
servers is $200$. Furthermore, the performance of the scheme based on
R10 codes is close to that of an ideal rateless code.

\bibliographystyle{IEEEtran}
\bibliography{manuscript}{}

\end{document}